\documentclass{ws-procs975x65}
\def\beq{\begin{equation}}
\def\eeq{\end{equation}}
\begin{document}

\title{Possible formation of lowly luminous highly magnetized white dwarfs by accretion leading to SGRs/AXPs}
\author{B. Mukhopadhyay$^{1*}$, M. Bhattacharya$^2$, A. R. Rao$^3$, S. Mukerjee$^1$, U. Das$^4$}

\address{1. Indian Institute of Science, Bangalore 560012, India\\
$^*$E-mail: bm@iisc.ac.in\\
%www.iisc.ac.in\\
2. University of Texas, Austin, USA\\
3. Tata Institute of Fundamental Research, Mumbai, India\\
4. University of Colorado, Boulder, USA
}

%\author{Tushar Mondal} 
%
%\address{Department of Physics, Indian Institute of Science, \\
%Bangalore 560012, India\\
%E-mail: mtushar@iisc.ac.in}

\begin{abstract}
We sketch a possible evolutionary scenario by which a highly magnetized super-Chandrasekhar white dwarf could be formed
by accretion on to a commonly observed magnetized white dwarf. This is an exploratory study, when the physics in 
cataclysmic variables (CVs) is
very rich and complex. Based on this, we also explore the possibility that the white dwarf pulsar AR Sco acquired its
high spin and magnetic field due to repeated episodes of accretion and spin-down. We show that strong magnetic field dramatically decreases luminosity of highly magnetized white dwarf (B-WD), letting them below the current detection limit. The repetition of this cycle can eventually lead to a B-WD, recently postulated to be the reason for over-luminous type Ia supernovae. A spinning B-WD could also be an ideal source for continuous gravitational radiation and soft gamma-ray repeaters (SGRs) and anomalous X-ray pulsars (AXPs). SGRs/AXPs are generally believed to be highly magnetized,
but observationally not confirmed yet, neutron stars. Invoking B-WDs does not require the magnetic field
to be as high as for neutron star based model, however reproducing other observed properties intact.
\end{abstract}


\bodymatter

%%%%%%%%%%%%%%%%% now a standard article style for the most part

\section{Introduction}

Several independent observations repeatedly argued in recent past for the existence of highly magnetized white dwarfs (B-WDs).
Examples are overluminous type Ia supernovae \cite{how,dmprl}, white dwarf pulsars \cite{arsco,mrmnras} etc. Also  
soft gamma-ray repeaters (SGRs) and anomalous X-ray pulsars (AXPs) could be explained as B-WDs
\cite{pac,usov,mrjcap}, while they are generally believed to be highly magnetized neutron stars \cite{magnetar} without 
however any direct detection of underlying required high surface field $B_s\sim 10^{15}$ G. Interestingly, explaining 
SGR/AXP by a magnetized white dwarf requires a lower $B_s\lesssim 10^{12}$ G, which may however 
correspond to central field $\gtrsim 10^{14}$ G. Nevertheless, the origin of such fields in a white dwarf remains 
a question, when the observed confirmed surface field is $\lesssim 10^9$ G.

Here we explore a possible evolution of a conventionally observed magnetized white dwarf to a B-WD
by accretion, which may pass through a phase exhibiting currently observed AR Sco. This is an exploratory study, 
and the present venture is based more on an idealized situation, when the 
physics in accreting white dwarfs, i.e. cataclysmic variables (CVs), is very rich and complex. We also show, based 
on some assumption, that the thermal luminosity of such a B-WD could be very small, below their current
detection limit. However, due to high field and rotation, their spin-down luminosity could be quite high. Hence, they
could exhibit SGRs/AXPs.

\section{Accretion induced evolution}

The detailed investigation of the accretion induced evolution 
faces several difficulties including nova eruptions (hence nonsteady increase of mass) and the eruption and ejection of 
accumulated shells. Nevertheless, the discovery of AR Sco, which is a fast rotating
magnetized white dwarf, argues for the possibility of episodic increase of mass in a CV. 
Hence, we sketch a tentative evolutionary scenario with repeated episodes of
accretion phase leading to the high magnetic field via flux freezing 
and spin-power phase decreasing field. Eventually this mechanism can plausibly lead to a B-WD.
Note that there are already observational evidences for transitions between spin-power and
accretion-power phases in a binary millisecond pulsar \cite{papi}.
The conservation laws controlling the accretion-power phase
%linear and angular momenta conservation and conservation of
%magnetic flux, 
around the stellar surface of radius $R$ and mass $M$, which could be inner edge of accretion disk,
%depending on the field strength, 
are given by
\begin{eqnarray}
l\Omega(t)^2 R(t)= \frac{GM(t)}{R(t)^2},\,\,
I(t)\Omega(t)={\rm constant},\,\,
B_s(t)R(t)^2={\rm constant},
\label{conv}
\end{eqnarray}
where $l$ takes care of inequality due to dominance of gravitational force
over the centrifugal force in general, $I$ is the moment of inertia of star and
$\Omega$ the angular velocity of the star which includes the additional contribution
acquired due to accretion as well. Solving the conservation laws given by equation (\ref{conv}) 
simultaneously, we obtain
the time evolution of radius (or mass), magnetic field and angular velocity during accretion.
Accretion stops when
\begin{eqnarray} 
-\frac{GM}{R^2}=\frac{1}{\rho}\frac{d}{dr}\left(\frac{B^2}{8\pi}\right)|_{r=R}
\sim -\frac{B_s^2}{8\pi R\rho},
\end{eqnarray}
where $\rho$ is the density of inner edge of disk.

For a dipolar fixed field, $\dot{\Omega}\propto\Omega^3$ \cite{mrjcap}, where over-dot implies time
derivative. Generalizing for the present purpose it becomes 
$\dot{\Omega}=k\Omega^n$ with $k$ being constant. Therefore,
during the phase of spin-power pulsar (when accretion inhibits),
the time evolution of angular velocity and surface magnetic field may be given by
\begin{eqnarray}
\label{omd}
\Omega=\left[\Omega_0^{1-n}-k(1-n)(t-t_0)\right]^{\frac{1}{1-n}},\,\,
%\end{eqnarray}
%\begin{eqnarray}
B_s=\sqrt{\frac{5c^3Ik\Omega^{n-m}}{R^6\sin^2\alpha}},
%\label{bd}
\end{eqnarray}
where $\Omega_0$ is the angular velocity when accretion just stops at the beginning of spin-power phase
at time $t=t_0$,
$k$ is fixed to constrain $B_s$ at the beginning of first spin-powered phase,
which is determined from the field evolution in the preceding accretion-power phase, $\alpha$ is the angle between magnetic and spin axes.
Note that $n=m=3$ corresponds to dipole field. 
%hence $m$ represents the deviation from dipolar field particularly for $n=3$.

\begin{figure}[h]
\begin{center}
\vskip-1.0cm
\includegraphics[width=3.3in]{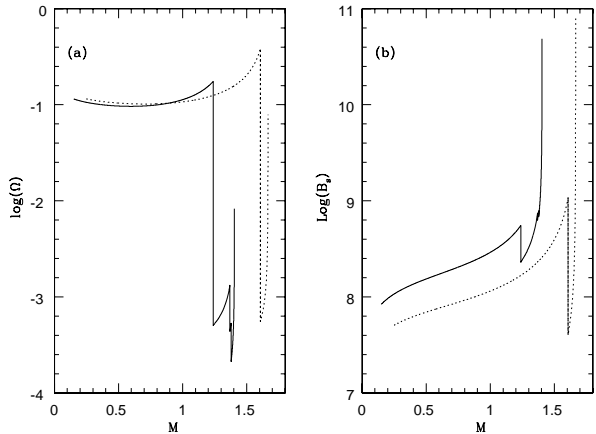}
\end{center}
\vskip-1.5cm
\caption{
Time evolution of (a) angular velocity in s$^{-1}$, (b) magnetic field
in G, as functions of mass in units of solar mass. The solid curves correspond
to $n=3$, $m=2.7$, $\rho=0.05$ gm cm$^{-3}$, $l=1.5$ and dotted
ones to $n=3$, $m=2$, $\rho=0.1$ gm cm$^{-3}$, $l=2.5$;
$k=10^{-14}$ CGS, $\dot{M}=10^{-8}M_\odot$Yr$^{-1}$,
$\alpha=10^o$ and $R=10^4$ km at $t=0$. This is reproduced from a previous work \cite{mrjcap}.
}
\label{timev}
\end{figure}

Figure \ref{timev} shows a couple of representative possible evolutions of angular
velocity and magnetic field with mass (time). 
%$\dot{M}$ in all the accretion-powered phases is chosen to be $10^{-8}M_\odot$Yr$^{-1}$,
%which is slightly higher than that of a typical
%intermediate polar, but an order of magnitude lower $\dot{M}$ would also suffice our purpose.
%Other parameters, mentioned in the Fig. \ref{timev} caption, are some of their typical representative values.
It is seen that initial larger $\Omega$
with accretion drops significantly during spin-power phase (when accretion stops and
hence no change of mass), followed by its increasing phase. 
Similar trend is seen in $B_s$ profiles
with a sharp increasing trend (with value $\sim 10^{11}$ G) at the last
cycle leading to the increase of $B_c$ as well, forming a B-WD. At the end of evolution,
it could be left out as a super-Chandrasekhar B-WD and/or a SGR/AXP candidate with a higher spin
frequency. Of course, in reality they may
depend on many other factors and the current picture does not match
exactly with what is expected in AR Sco itself.

\section{Luminosity}

With the increase of mass, the radius of white dwarfs, hence B-WDs, becomes very small \cite{dmr,mrmnras}.
Indeed, the increase of magnetic field is due to decreasing radius via flux-freezing.
Now due to smaller radius, UV-luminosity of B-WDs turns out to be very small if the 
surface temperature is same as their nonmagnetic counterpart \cite{mrjcap}. However more
interestingly, from the conservation of energy, it is expected that the presence of 
strong magnetic field enforces decreasing thermal energy and hence lowering luminosity
in stable equilibrium.

\begin{figure}[h]
\begin{center}
\includegraphics[width=2.2in]{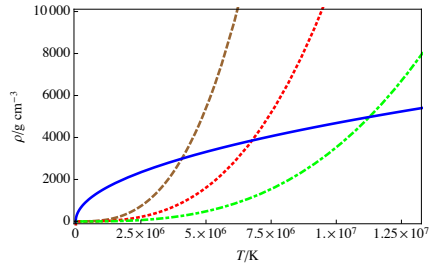}
\end{center}
\caption{Variation of density with temperature for $B\equiv (B_{\rm{s}},B_0) = (10^{12}\,\rm{ G},10^{14}\,\rm{G})$ and $L=10^{-5}L_{\odot}$ (dashed line), $10^{-4}L_{\odot}$ (dotted line) and $10^{-3}L_{\odot}$ (dot-dashed line).  
%$\rho_{*}$ and $T_{*}$ are obtained from the intersection of the $\rho-T$ profiles with equation (\ref{eqn_rhoT_B}) (solid line). 
The solid line represents equation (\ref{int}).
This is reproduced from a previous work \cite{mukul}.
}
\label{trhoint}
\end{figure}

Combining the magnetostatic and photon diffusion equations in the presence of magnetic 
field but ignoring tension, we obtain 
\begin{eqnarray}
\frac{d}{dT}\left(P+P_B\right)=\frac{4ac}{3}\frac{4\pi GM}{L}\frac{T^3}{\kappa},
\label{hydph}
\end{eqnarray}
which we solve to obtain the envelop properties. Here $P$ is the matter pressure, 
$P_B$ the magnetic pressure, 
%$\rho$ the matter density, 
$\kappa$ the opacity, 
$T$ the temperature, $a$ the radiation constant, 
%$c$ the speed of light, $G$ the Newton's gravitational constant, 
$M$ the mass of white dwarf within the core radius 
$r$, which is practically the whole mass of white dwarf because the envelop is very thin, 
and $L$ is the luminosity. For the strong field considered here, the radiative opacity 
variation with $B$ can be modelled similarly to neutron stars as 
$\kappa=\kappa_B\approx 5.5\times 10^{31}\rho T^{-1.5}B^{-2}$ cm$^2$g$^{-1}$ \cite{py01}.
We use a field profile proposed earlier for neutron stars \cite{bandprl} to enumerate
the field magnitude at a given density (radius), irrespective of other complicated effects,
given by
\begin{eqnarray}
B\left(\frac{\rho}{\rho_0}\right)=B_s+B_0\left[1-\exp\left(-\eta\left(\frac{\rho}{\rho_0}\right)^\gamma\right)\right],
\label{bprof}
\end{eqnarray}
where $B_0$ (similar to central field) is a parameter with the dimension of $B$, other parameters 
are set as $\eta=0.1$, $\gamma=0.9$, $\rho_0=10^9$ g cm$^{-3}$ for all the calculations. Further
equating the electron pressure for the non-relativistic electrons on both sides of the core-envelop
interface gives
\begin{eqnarray}
\rho_*(B_*)\approx1.482\times 10^{-12}~T_*^{1/2}B_s
\label{int}
\end{eqnarray}
at interface.
Now we solve equation (\ref{hydph}) along with the photon diffusion equation
\begin{eqnarray}
\frac{dT}{dr}=-\frac{3}{4ac}\frac{\kappa(\rho+\rho_B)}{T^3}\frac{L}{4\pi r^2},
\label{phdiff}
\end{eqnarray}
with boundary conditions $\rho(T_s)=10^{-10}$ g cm$^{-3}$,
$r(T_s)=R=5000$ km and $M=M_\odot$, where $T_s$ is the surface temperature, and obtain $\rho-T$ and $r-T$ profiles.
Further $T_*$ and $\rho_*$ can be obtained by solving for the $\rho-T$ profile  
along with equation (\ref{int}), as shown in Fig. \ref{trhoint}, and knowing 
$T_*$, we can obtain $r_*$ from the $r-T$ profile. We see that interface 
moves inwards ($r_*$ decreases) with increasing $B$ and $L$. 
%Further, the temperature fall rate close to the surface increases with increasing $L$ and decreasing $B$. 
But $\rho_*$ increases with increasing $L$ and/or $B$,
as $\rho_*\propto T_*^{1/2}B$. 

With above benchmarking, we now explore, based energy conservation, if the luminosity
of a magnetized white dwarf or B-WD changes. For $B=0$ and $L=10^{-5}L_\odot$, we have
$r_*=0.9978R$, $\rho_*=170.7$ g cm$^{-3}$ and $T_*=2.332\times 10^6$ K.
Using the same boundary condition as described above, we now solve 
equations (\ref{hydph}) and (\ref{phdiff}) with $B\neq 0$, but vary $L$ in order to fix
$r_*=0.9978R$. We find interestingly that $L$ decreases for $B\neq 0$, as shown in Table~1. Physically
this corresponds to increasing $B$, and thence magnetic energy, is compensated by
decreasing thermal energy (decreasing $T_*$) and thence $L$, when total energy is conserved. Similarly,
increasing $B$ may be compensated by decreasing gravitational energy (decreasing $r_*$).
In either of the cases, $L$ decreases. 

\begin{table}
\begin{center}
%\small
%\caption{\small Table~1: Variation of luminosity with magnetic field for fixed $r_{*}=0.9978R$}
{Table~1: Variation of luminosity with magnetic field for fixed $r_{*}=0.9978R$}
\begin{tabular}{cccccccccccccccccccccc}
\hline
\hline
\centering
$B/\rm{G}=(B_{\rm{s}}/G,B_0/G)\,$ & $L/L_{\odot}$ &  $T_{*}/\rm{K}$ & $\rho_{*}/\rm{g\,cm^{-3}}$ & $T_{\rm{s}}/\rm{K}$ \\ \hline
$(0, 0)$                 & $1.00\times10^{-5}$ & $2.332\times10^{6}$ & $1.707\times10^{2}$ & $3.85\times10^{3}$ \\ \hline
$(10^9,6\times10^{13})$ & $2.53\times10^{-7}$ & $4.901\times10^{5}$ & $1.037\times10^{0}$ &$1.53\times10^{3}$ \\ \hline
%$(2\times10^{9},4\times10^{13})$ & $2.07\times10^{-8}$ & $2.737\times10^{5}$ & $1.551\times10^{0}$ &$8.21\times10^{2}$ \\ \hline
$(5\times10^{9},2\times10^{13})$ & $3.96\times10^{-8}$ & $3.262\times10^{5}$ & $4.232\times10^{0}$ &$9.65\times10^{2}$ \\ \hline
$(10^{10},10^{13})$ & $1.02\times10^{-6}$ & $7.189\times10^{5}$ & $1.257\times10^{1}$ &$2.17\times10^{3}$ \\ \hline
%$(2\times10^{10},6\times10^{12})$ & $1.22\times10^{-6}$ & $7.616\times10^{5}$ & $2.587\times10^{1}$ &$2.27\times10^{3}$ \\ \hline
$(2\times10^{10},8\times10^{12})$ & $4.40\times10^{-9}$ & $2.063\times10^{5}$ & $1.346\times10^{1}$ &$5.57\times10^{2}$ \\ \hline
$(5\times10^{10},4\times10^{12})$ & $2.59\times10^{-8}$ & $3.185\times10^{5}$ & $4.182\times10^{1}$ &$8.68\times10^{2}$ \\ \hline
%$(10^{11},2\times10^{12})$ & $1.09\times10^{-6}$ & $7.721\times10^{5}$ & $1.302\times10^{2}$ &$2.21\times10^{3}$ \\ \hline
$(5\times10^{11},10^{12})$ & $2.93\times10^{-9}$ & $2.206\times10^{5}$ & $3.480\times10^{2}$ &$5.03\times10^{2}$ \\ \hline
\hline
\label{table5}
\end{tabular}
\end{center}
\end{table}

\section{SGRs/AXPs as B-WDs}

Paczynski \cite{pac} and Usov \cite{usov} independently proposed that 
SGRs and AXPs 
are moderately magnetized white dwarfs but following Chandrasekhar's
mass-radius relation \cite{chandra}. Many features of SGRs/AXPs
are explained by their model at relatively lower magnetic fields,
while the more popular magnetar model \cite{magnetar} requires field 
$\gtrsim 10^{15}$ G, which is not observationally well established yet. 
Nevertheless, such a white dwarf based model suffers from a
deep upper limit on the optical counterparts of some  AXPs/SGRs, 
e.g. SGR 0418+5729, due to their larger moment of inertia. 

Now B-WDs established here could be quite smaller in size and hence have
smaller moment of inertia. Therefore, their optical counterparts,
with very low UV-luminosities, are quite in accordance with observation. 
Hence, the idea of B-WD brings a new scope of explain SGRs/AXPs at smaller magnetic fields,
which are observationally inferable, compared to highly magnetized magnetar 
model. For details see the work by Mukhopadhyay \& Rao \cite{mrjcap}.

\section{Continuous Gravitational Radiation}

Due to smaller size compared to their regular counterpart, B-WDs rotate relatively faster.
Now if the rotation and magnetic axes are misaligned, they serve as good candidates
for continuous gravitational radiation 
due to their quadrupole moment, characterized by the amplitude \cite{palomba}
\begin{equation}
h_+(t)=\frac{h_0}{2}(1+\cos^2\alpha_0)\cos\Phi(t),\,\,h_{\times}(t)=h_0\cos\alpha_0\sin\Phi(t),\,\,h_0=\frac{4\pi^2G I_{zz}\epsilon}{c^4 P_s^2 D},
\label{h+x}
\end{equation}
where $\alpha_0$ is the inclination of the star's rotation axis with respect to the observer,
%\begin{equation}
%h_0=\frac{4\pi^2G}{c^4}\frac{I_{zz}\epsilon}{P_s^2 D},
%\label{ho}
%\end{equation}
$\Phi(t)$ is the signal phase function, $\epsilon$ amounts the ellipticity of the star,
$I_{zz}$ is the moment of inertial about z-axis, and
$D$ is the distance between the star and detector.

A B-WD of mass $\sim 2M_\odot$,
polar radius $\sim 700$ km, spin period $P_s\sim 1$ s \cite{sathya}, $\epsilon\sim 5\times 10^{-4}$ and
$D\sim 100$ pc 
would produce $h_0\sim 10^{-22}$, which is within the sensitivity
of the Einstein@Home search for early Laser Interferometer Gravitational Wave Observatory
(LIGO) S5 data \cite{palomba}.
%to be detected by LIGO/VIRGO from typical neutron stars at about $10$ kpc away (\citealt{palomba}).
However, DECIGO/BBO would give a firm confirmation of their gravitational wave 
because they are more sensitive in their frequency range.
In fact, if the polar radius is $\sim 2000$ km with $P_s\sim 10$ s and other parameters intact,
DECIGO/BBO can detect it with $h_0\sim 10^{-23}$.
%LISA could probe it with $h_0\sim 10^{-24}$.
Nevertheless, (highly) magnetized rotating white dwarfs approaching B-WDs are expected to 
be common and such white dwarfs of radius $\sim 7000$ km, $P_s\sim 20$ sec and $D\sim 10$ pc
could produce $h_0 \gtrsim 10^{-22}$ which is detectable by LISA.

\section{Summary}

The idea of B-WD has been proposed early this decade, mainly to explain
observed peculiar type Ia supernovae inferring super-Chandrasekhar 
progenitor mass. Lately it has been found with various other applications,
e.g. SGRs/AXPs, white dwarf pulsars like AR Sco, continuous gravitational 
wave etc. Here we have attempted to sketch a plausible evolution scenario 
to explain the formation of such a highly magnetized, smaller size white dwarf.
In our simplistic picture, ignoring many complicated CV features, 
we are able to show that a commonly observed magnetized white dwarf 
could be evolved to a B-WD via accretion. Hence, the existence of highly magnetized,
rotating, smaller white dwarfs is quite plausible.

\end{document}